\documentclass{kluwer}    % Specifies the document style.
\usepackage{graphicx}
\begin{document}
\begin{article}
\begin{opening}
\title{Self-consistent Modelling of the interstellar medium\thanks{XMM-Newton is
an ESA Science Mission with instruments and contributions directly
funded by ESA Member states and the USA (NASA). The XMM-Newton
project is supported by the Bundesministerium für Bildung und
Forschung/Deutsches Zentrum f\"ur Luft- und Raumfahrt (BMBF/DLR),
the Max-Planck-Gesellschaft and the Heidenhain-Stiftung.}}
\author{Dieter \surname{Breitschwerdt}}
\runningauthor{Dieter Breitschwerdt} \runningtitle{Self-consistent
modelling} \institute{Max-Planck-Institut f\"ur extraterrestrische
Physik, Postfach 1312, D-85741 Garching, Germany; Email:
\texttt{breitsch@mpe.mpg.de} }
\date{November 30, 2002}

\begin{abstract}
The dynamical evolution of hot optically thin plasmas in the ISM
crucially depends on the heating and cooling processes. It is
essential to realize that all physical processes that contribute
operate on different time scales. In particular {\em detailed
balancing} is often violated since the statistically {\em inverse}
process of e.g.\ collisional ionization is recombination of an ion
with two electrons, which as a three-body collision is usually
dominated by radiative recombination, causing a departure from
collisional ionization equilibrium. On top of these differences in
atomic time scales, hot plasmas are often in a dynamical state,
thereby introducing another time scale, which can
be the shortest one.\\
The non-equilibrium effects will be illustrated and discussed in
the case of galactic outflows. It will be shown, that spectral
analyses of X-ray data of edge-on galaxies show a clear signature
in the form of ``multi-temperature'' halos, which can most
naturally be explained by the ``freezing-in'' of highly ionized
species in the outflow, which contribute to the overall spectrum
by {\em delayed recombination}. This naturally leads to a
non-equilibrium cooling function, which modifies the dynamics,
which in turn changes the plasma densities and thermal energy
budget, thus feeding back on the ionization structure. Therefore
{\em self-consistent} modelling is needed.

\end{abstract}

\end{opening}

\section{Introduction}
\label{intro}
Interstellar plasmas are subject to numerous dynamical and thermal
processes, which are essentially driven by star formation activity
in galaxies. Matter and radiation are continuously exchanged
between the interstellar medium (ISM) and stars. This can loosely
be described as a matter cycle. Star formation induced by
gravitational instabilities is only possible if both transport of
angular momentum, mass and energy work efficiently, with radiative
cooling playing a key r\^ole both in the formation of interstellar
clouds and protostars. During its main sequence lifetime a star
interacts with the ISM by its constant photon output, by which
especially Lyman-continuum photons can inject a substantial amount
of energy creating HII regions. Moreover mass loss in the form of
stellar winds gives rise to hot shock heated extended bubbles due
to a mechanical wind luminosity of the order of $10^{36}$ erg/s,
corresponding to about 30\% of the energy of a type~II supernova
(SN) integrated over the stellar main sequence life time. However,
the most efficient energy and momentum input into the ISM is by SN
explosions, both randomly or in concert thus generating
superbubbles. Shock velocities are initially of the order of
$10^3$ km/s, and therefore from simple considerations of
Rankine-Hugoniot conditions for strong shocks and adiabatic
coefficient $\gamma = 5/3$, the post shock temperature obeys the
simple relation: $ T_{sh} \approx {3\over 16} (\mu \bar m/k_B)
v_8^2 \approx 2.7 \times 10^7 \mu v_8^2 \, {\rm K}$, with $\mu$,
$\bar m$, $k_B$ and $v_8$ being the mean molecular weight, mean
particle mass, Boltzmann's constant and the upstream shock
velocity (in the shock rest frame) in units of 1000 km/s,
respectively. Thus radiative cooling will be by thermal
bremsstrahlung, recombination lines as well as atomic line
transitions, detectable mainly in soft X-rays. It should be
emphasized however, that the bulk of the energy is lost by
adiabatic expansion of the supernova remnant (SNR) or superbubble
(SB). This can be attributed to the fact that there exists a huge
overpressure with respect to the ambient medium, acting like a
piston, and that the radiative cooling time of low density bubbles
is usually much larger than the timescale by which they merge with
the ISM. As a rough estimate, we can compare the energy loss rate
per unit volume with the $p\, dV$-work done by the expanding
bubble: $\Xi=n^2 \Lambda(T) t/(2 n k_B T) \approx 1.1 \times 10^7
(n/T) t_6 < 1$, as long as $T \gtrsim 10^6$ K, using
$\Lambda(T)\lesssim 10^{-22} \, {\rm erg} \, {\rm cm}^3/{\rm s}$
and $t_6$ being the SB lifetime in units of Myr, where $T$ and $n$
are the plasma temperature and density, respectively. Note that
during the adiabatic stage of evolution, similarity solutions may
be applied, thus $n \propto t^{\alpha}$, with $\alpha = -4/5$ for
SBs and $\alpha = -6/5$ for SNRs. For $T \sim \mathrm{const}. $
during the energy driven phase this implies $\Xi \propto t^{1/5}$
for SBs and $\Xi \propto t^{-1/5}$ for SNRs. Thus if $\Xi(t) \ll
1$ initially, it will be so later on due to the weak time
dependence of $\Xi$. For $n<10^{-3} \, {\rm cm}^{-3}$, $T > 10^6$
K, as it is expected for superbubbles, this will definitely hold,
except, maybe, for bubbles in very dense regions.

%The shocked interstellar plasma, in contrast to the shocked wind
%and ejecta gas, has a higher density if the shock expands into a
%non-rarefied ISM, and thus is subject to strong cooling, leading
%to dense partially neutral shells. Compressible media can easily
%undergo transitions between stable phases due to effective heating
%and cooling mechanisms. Thermal instabilities occur when $d(\log
%T)/d(\log n) < -1$.

On a larger scale, where star forming regions (SFRs) are
clustered, the combined pressure of winds and SN explosions can
drive a galactic fountain or wind. Such a flow is again
characterized by strong adiabatic and radiative cooling. Most
prominent examples are starburst galaxies, from which so-called
``superwinds'' emanate (e.g. Heckman et al., 1990). Common to all
examples mentioned here is a \emph{coupling} between the
\emph{thermal} and the \emph{dynamical} evolution, which will be
described and discussed in the following sections.

\section{Non-equilibrium cooling of interstellar plasmas}
\label{NEI}
Interstellar cooling is on a fundamental level a complicated
process. The energy loss rate per unit volume, expressed as the
cooling function, $\Lambda(T,Z)$, when normalized to the electron
and ion densities, not only depends on temperature $T$ and the
elemental abundances $Z$ of the fluid, but in general also on
the distribution of ionization stages. The assumption of
collisional ionization equilibrium (CIE) circumvents this problem.
It is assumed that the rate of ionization by inelastic collions
between electrons and ions is exactly balanced by the rate of
recombinations. While the former is a cooling process the latter
as a statistically inverse process must consequently be a heating
process. Since the energy released has to be transferred to a
third particle, thereby increasing the thermal energy of the
plasma, such three-body interactions are extremely improbable in a
tenuous, hot optically thin plasma like in a bubble or a
superwind. Therefore it is commonly taken that \emph{radiative
recombination} is dominating; however, this is also a cooling
process to the plasma. Thus a hot plasma, left to itself will
eventually depart from equilibrium (s. Shapiro \& Moore, 1976;
Schmutzler \& Tscharnuter, 1993).

\begin{figure}[!t]
 \includegraphics[width=0.7\textwidth,clip]{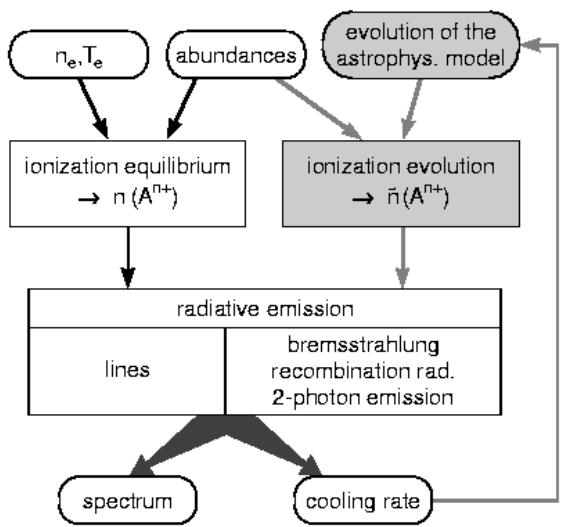}
\caption[]{Flow chart of plasma emission model (s.~B\"ohringer,
1997). White boxes refer to equilibrium (CIE), grey boxes to
additions necessary for non-equilibrium (NEI) models. $A^{n+}$
denotes atomic species $A$ ionized n-times.} \label{chart}
\end{figure}

The situation can be more dramatic, if the plasma undergoes
compression like e.g. in a shock wave, or expansion like in a SB,
fountain or wind. The first case gives rise to so-called
\emph{underionized} plasmas: the electrons downstream of a
collisionless shock are heated up very efficiently by randomizing
their kinetic energy via turbulent electromagnetic fields, and on
a somewhat larger timescale also the ions. The ionization process
on the other hand depends on the cross sections and thus on atomic
timescales, which are much larger. Therefore the ionization will
lag behind, and when observed spectroscopically, the plasma will
appear cooler than it actually is (see e.g. Cox \& Anderson,
1982). The second case has the opposite characteristics, and is
known as an \emph{overionized} plasma: fast adiabatic expansion,
like e.g. in an expanding wind, which occurs on much shorter
timescales than radiative recombination, makes the plasma appear
hotter to the observer due to \emph{delayed recombination} (see
Breitschwerdt \& Schmutzler, 1994). It has been shown that in
these non-equilibrium ionization (NEI) cases the dynamics of the
plasma and its thermal evolution are closely intertwined. The
dynamics changes the density and temperature, which
in turn changes the ionization structure, which determines the
cooling function, which again modifies the dynamics. Thus a
self-consistent description is necessary.
\begin{figure}[!t]
\mbox{%
 \includegraphics[bb=155 307 433 526,width=0.49\textwidth,clip]{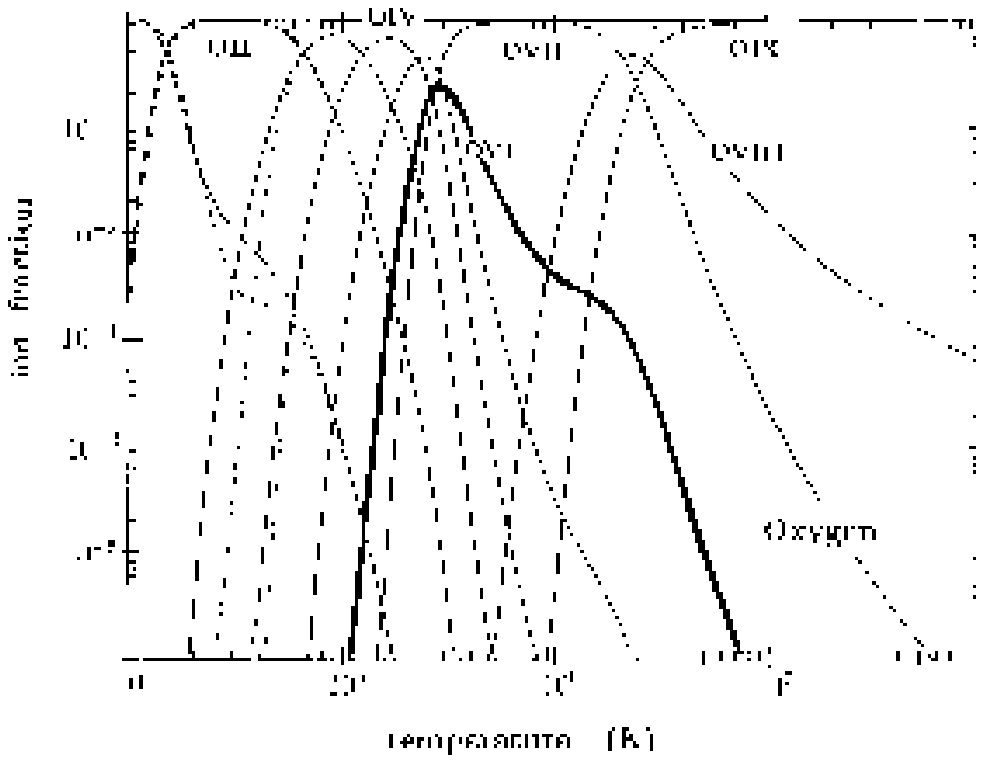}
 \hspace*{0.5mm}
 \includegraphics[width=0.49\textwidth,clip]{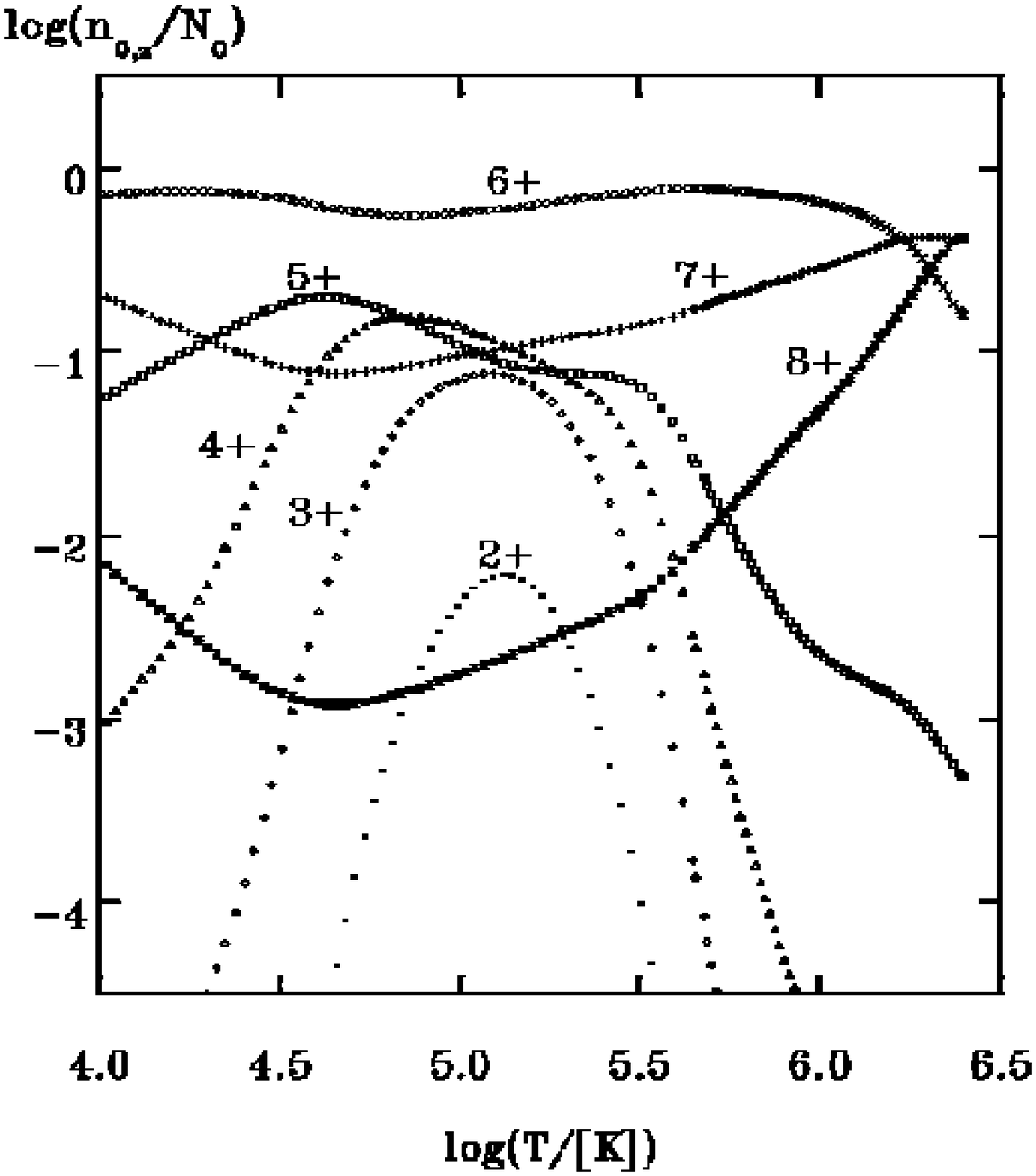}
}
%
%\vspace{6pc}
\caption[]{\textbf{Left:} Ionization fractions for oxygen as a
function of temperature (taken from B\"ohringer and Hensler, 1989)
for a plasma in collisional ionization equilibrium (CIE).
\textbf{Right:} Ionization fractions for oxygen as a function of
temperature for a plasma in non-equilibrium (NEI) taken from
Breitschwerdt and Schmutzler (1999). }
\label{ion_ox}
\end{figure}
Under CIE assumptions, the distribution of ionization stages is a
sensitive function of temperature and thus the measurement of line
ratios (e.g. NV/OVI) leads to a reliable temperature
determination; however, in the case of NEI this is no longer valid
(s.~Fig.~\ref{ion_ox}). In essence the plasma keeps a memory of
its origin, and due to fast adiabatic cooling a particular ion can
be present over a wide range of temperatures. Therefore when
modelling the emissivity and the spectrum of a plasma, the
time-dependent evolution of the ionization stages has to be
followed closely along with the evolution of the astrophysical
model (s.~Fig.~\ref{chart}).
Before turning to some detailed examples, some caveats for using
plasma models have to be pointed out: although there exists a large
data base of atomic cross sections for the various processes
included, such as collisional and auto-ionization,
photo-ionization, radiative and dielectronic recombination,
two-photon emission, Auger effect, charge exchange reactions, just
to name the most important ones, there are still quantitative
discrepancies in plasma codes on the market (cf. Masai, 1997),
particularly for heavy elements like iron around 1 keV (up to a
factor of 2 in emissivity). Moreover, when fitting observed X-ray
spectra it is strictly
speaking wrong to substitute a CIE fit model (e.g.\ RS, MEKAL
etc.) by a ``non-equilibrium model'' that has just one or more
free parameters, like e.g. the ionization parameter $I = n_e t$,
simply because it is not calculated self-consistently with the
dynamical evolution. The evolution of the astrophysical model is a
\emph{necessary} although not sufficient input. It cannot even be
claimed that such a ``NEI fit model´´ is closer to the truth than
the CIE model, if it cannot be physically motivated; it just fits
better due to a larger number of free parameters.

\section{Spectral signatures of galactic outflows}
\label{outflow}
\begin{figure}[!t]
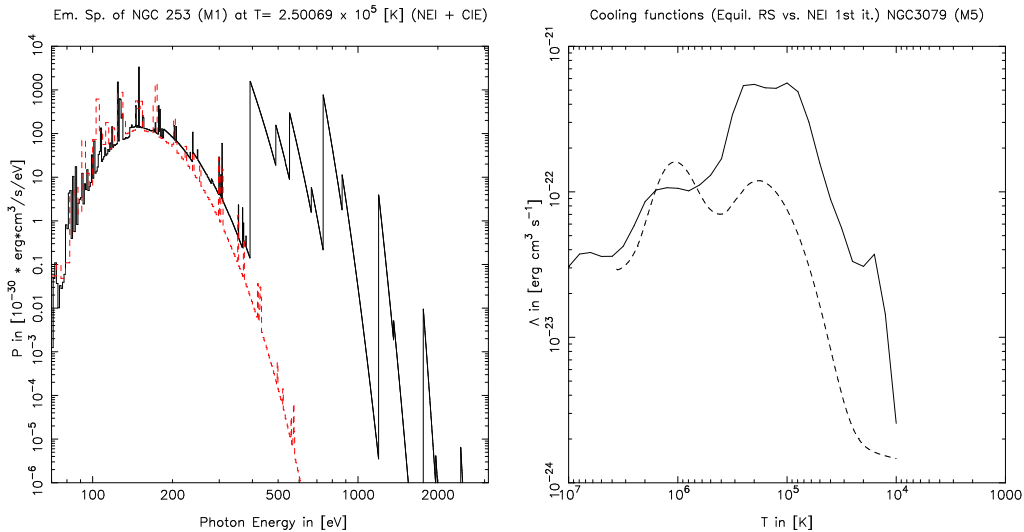

\label{n253-spec}
\mbox{%
 \includegraphics[width=0.6\textwidth,bbllx=35pt,bblly=20pt,bburx=585pt,%
                   bbury=535pt,angle=-90,clip]{breitschwerdt_fig6.ps}
 \hspace*{0.5mm}
 \includegraphics[width=0.6\textwidth,bbllx=35pt,bblly=25pt,bburx=585pt,%
                   bbury=543pt,angle=-90,clip]{breitschwerdt_fig7.ps}
}
%\vspace{6pc}
\caption[]{\textbf{Left:} Comparison of a CIE and a NEI emission
spectrum resulting from a simulated outflow of the starburst
galaxy NGC\,253 at a distance (and time) when the temperature has
cooled down $2.5 \times 10^5$ K. The characteristic saw-tooth
structure is due to recombination lines, here of SiXII and FeXVII
(two main peaks, with energy increasing). \textbf{Right:}
Comparison of CIE (solid curve) cooling function $\Lambda$ and the
one resulting from a self-consistent evolution (dashed curve) of
ionization stages for the successful spectral fit model of the
halo emission of NGC\,3079.

  }
\end{figure}

In the following it will be demonstrated that NEI cooling leaves a
distinct imprint on the plasma, which can already be seen by
current soft X-ray observations, even at modest spectral
resolutions. The reason being the temperature sensitivity of CIE
spectra and the existence of diagnostic ions, such as lithium and
helium like oxygen and iron L shell transitions.
\begin{figure}[!t]
\label{n253-img}
\mbox{%
% \includegraphics[width=0.5\textwidth,bbllx=89pt,bblly=299pt,bburx=527pt,%
%                   bbury=738pt,clip]{breitschwerdt_fig4.ps}
 \includegraphics[width=0.5\textwidth,clip]{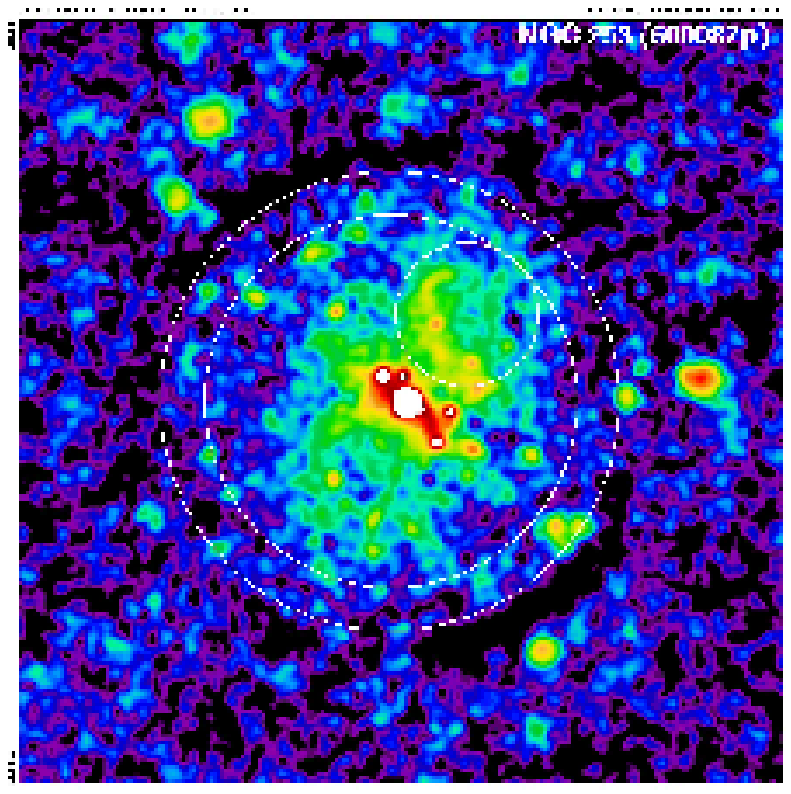}
 \hspace*{0.5mm}
 \includegraphics[width=0.7\textwidth,angle=90,bbllx=560pt,bblly=55pt,bburx=74pt,%
                   bbury=400pt,clip]{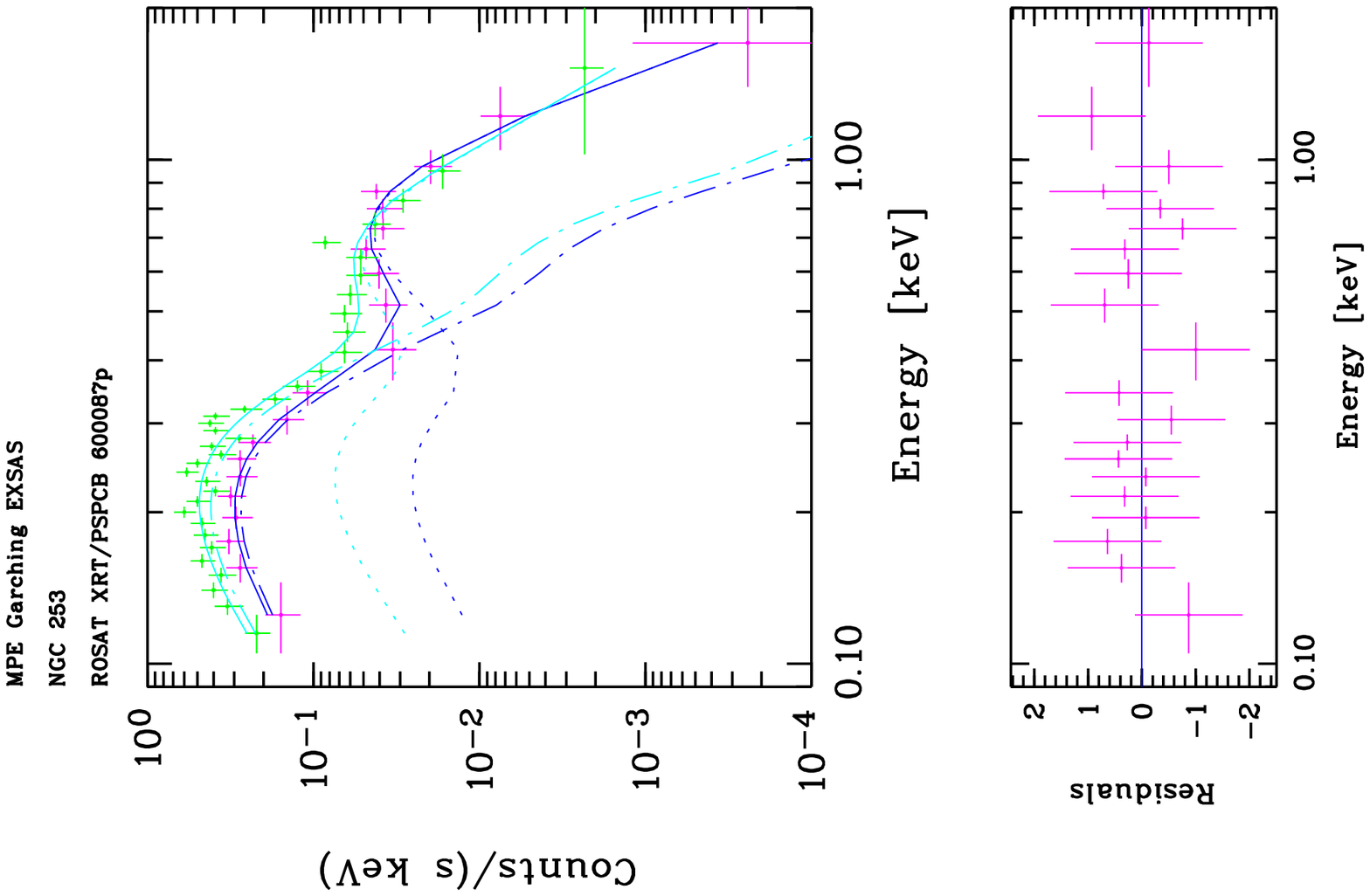}
}
%\vspace{6pc}
\caption[]{\textbf{Left:} ROSAT PSPC observation of the edge-on
starburst galaxy NGC\,253
  (north is up) in soft X-rays (0.1 - 2.4 keV) smoothed with a Gaussian
  filter; colour coding is from blue or dark (low) to red or bright (high intensity).
  The small circle
  shows the halo region that has been analyzed, while the outer
  ring represents the region that has been used for background subtraction.
\textbf{Right:} Comparison between data (red points with error
bars, lower curve) and self-consistent model M4
  (green points, upper curve). The solid, dashed, dash-dotted and dotted curves are
  plasma fit models (2 Raymond \& Smith (1977) CIE models) with solar abundances,
  respectively). There is also good
  {\em quantitative} agreement between data and model, when taking into account
  missing flux in the
  observations due to point source masking. %
  }
\end{figure}
We have analyzed 23 ksec of ROSAT PSPC archival data in the $0.1 -
2.4$ keV range of the nearby edge-on galaxy NGC\,253
(Breitschwerdt \& Freyberg, 2003) retaining 15.3 ksec of good
background subtracted data (s.~Fig.~\ref{n253-img}) and obtained
only a statistically acceptable fit by using \emph{two} CIE plasma
temperatures (applying the plasma code of Raymond \& Smith, 1977
(RS)); alternatively decreasing the elemental abundances to about
$0.05 Z_{\odot}$ was equally satisfactory. However, it seems
rather odd, that a SN heated and thus chemically enriched
superwind should have highly subsolar abundances (although this
has been claimed in the literature). On the other hand, we have
simulated the outflow by a galactic wind model (for details
s.~Breitschwerdt \& Schmutzler, 1999) calculating
self-consistently the full NEI plasma structure as discussed
before. The resulting spectrum, integrated over the field of view
(FOV) of the PSPC, binned into PSPC channels and folded through
the instrumental response, was both quantitatively and
qualitatively similar to the observed one
(s.~Fig.~\ref{n253-spec}). We have thus treated the
\emph{observed} and the \emph{simulated} spectra in exactly the
same manner. Moreover, spectral fitting of the \emph{model data}
was again satisfactory for 2 RS plasmas or abundances of $Z\approx
0.05 Z_{\odot}$. Hence it is fair to conclude that highly subsolar
abundances are an artefact of the fit procedure. Clearly
decreasing the abundances, washes out the prominent and
characteristic lines of a spectrum and therefore gives more
flexibility to accommodate all the temperature sensitive lines in
a CIE spectrum!
The boundary conditions of the outflow model reflect the SFR in
the underlying galactic disk, but are specified at a vertical
distance of $z=1$ kpc at the base of the halo. The flow geometry
is represented by a flux tube, following the transition from
planar close to the disk to spherically divergent flow at large
distances. The gravitational potential has been described by
(oblate) spheroids, including a bulge, a disk and a dark matter
halo component (to account for the flatness of the rotation
curve), scaled to observations. For the successful model, hot
plasma is injected at the base of the halo at a temperature $T =
2.5 \times 10^6$ K and a density $n= 7 \times 10^{-3}\, {\rm
cm}^{-3}$.
%% Give outflow base velocity and estimate mass loss rate
%
It is noteworthy that the models are fairly well constrained, even
by the PSPC spectral resolution. The intrinsic outflow spectra,
corresponding to disk parallel cuts through the halo at a given
distance from the galactic plane, are radically different from CIE
spectra (s.~Fig.~\ref{n253-spec}), characterized by strong
saw-tooth type recombination lines due to delayed recombination.
Note that both the elements and the ionization stages of the most
intense lines change as a function of time (or distance from the
disk), characterized by their recombination time scale. Another
marked difference between CIE and NEI spectra is the existence of
high energy X-ray photons in the overionized case (at around 1 keV
in Fig.~\ref{n253-spec}), although the temperature is already down
to $2.5 \times 10^5$ K. This occurs, because the plasma was once
hot enough to have generated these photons in the past: the plasma
retains a memory of its origin. The integrated spectrum, which is
folded through the detector response matrix for comparison with
the observed data (s.~Fig.~\ref{n253-img}), is the sum of such
intrinsic spectra over the FOV.
\begin{figure}[!t]
\label{n3079}
\mbox{%
 \includegraphics[width=0.49\textwidth,bbllx=1pt,bblly=1pt,bburx=428pt,%
                   bbury=444pt,clip]{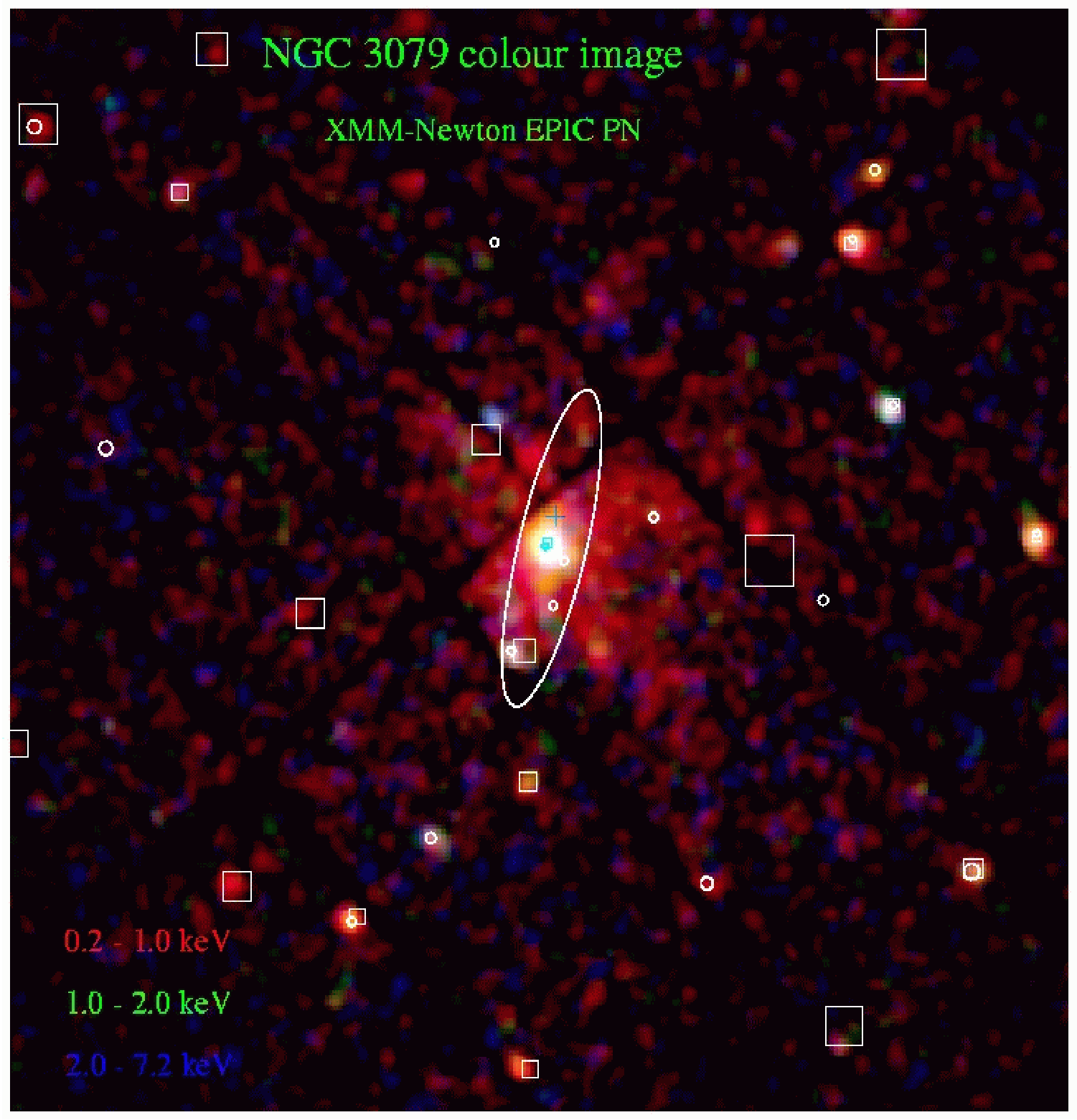}
 \hspace*{0.5mm}
% \includegraphics[width=0.49\textwidth,bbllx=100pt,bblly=206pt,bburx=473pt,%
%                   bbury=580pt,clip]{breitschwerdt_fig9.ps}
 \includegraphics[width=0.49\textwidth,clip]{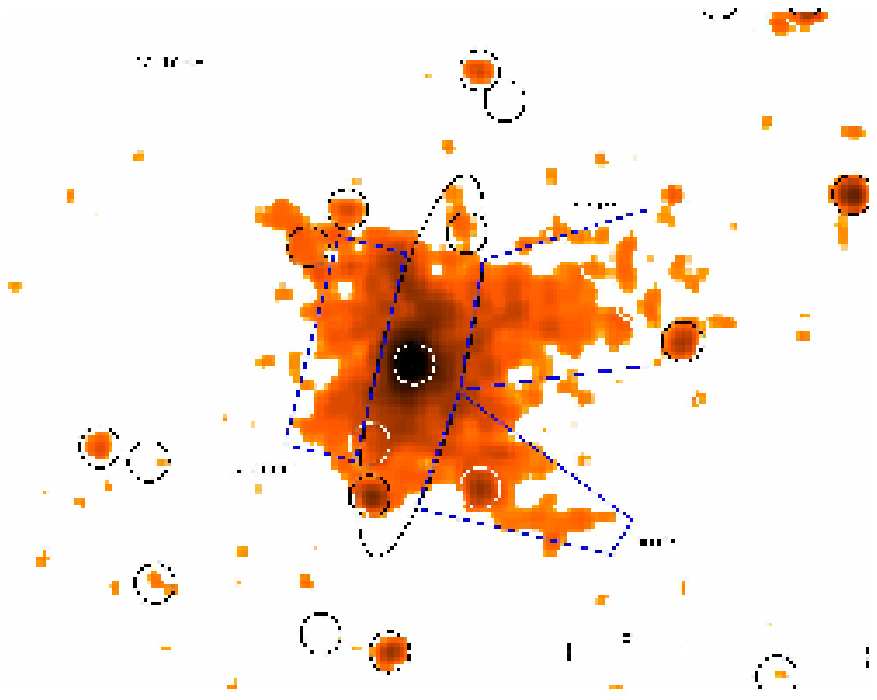}
}
%\vspace{6pc}
\caption[]{\textbf{Left:} RGB image of the extended X-ray halo of
  the nearby edge-on starburst galaxy NGC\,3079, as observed by the
  XMM-Newton EPIC pn camera.
  \textbf{Right:}  Smoothed image of the diffuse emission
  of NGC\,3079 in the 0.2 - 1.0 keV energy band
  (Breitschwerdt et al. 2003). Point sources are shown by
  circles. The $D_{25}$
  ellipse is indicated to show the optical extension of the disk.

  }
\end{figure}

Another striking example of an overionized outflow is the LINER
edge-on starburst galaxy NGC\,3079, which we have observed with
XMM-Newton for 25 ksec (Breitschwerdt et al. 2003). The image
taken with the EPIC pn camera shows an extended soft X-ray halo
(especially impressive in the $0.2 - 1$ keV band) of roughly the
same size as the $D_{25}$-ellipse, representing the optical
diameter of the disk (s.~Fig.~\ref{n3079}). It can be clearly seen
that apart from the nuclear outflow, there also exists an extended
component along the major and minor axes. Its spiky morphology is
probably related to the SFR in the underlying disk. We have
analyzed and fitted the EPIC pn spectrum of the NE spur
(Breitschwerdt et al., 2003).
\begin{figure}[!t]
\label{sbgal_fit}
\mbox{%
 \includegraphics[angle=-90,bb= 550 35 80 705,width=0.55\textwidth,clip]{breitschwerdt_fig10.ps}
 \hspace*{0.5mm}
 \includegraphics[width=0.44\textwidth,bbllx=60pt,bblly=50pt,bburx=500pt,%
                   bbury=440pt,clip]{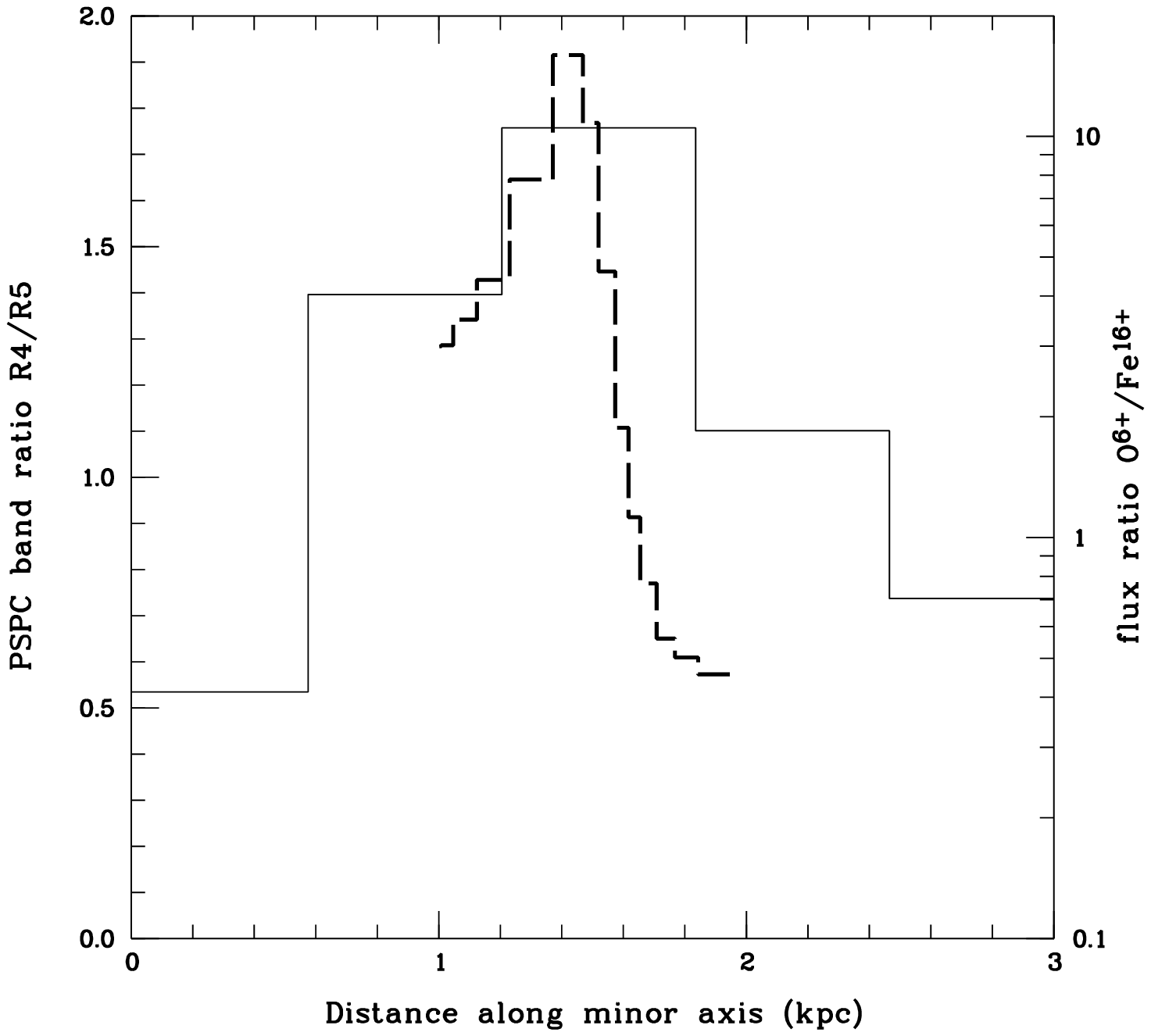}
}
%\vspace{6pc}
\caption[]{\textbf{Left:} Fit (solid line: $\chi^2 = 1.2$) of the
XMM-Newton Epic pn observed spectrum (crosses correspond to
  error bars) of the emission of NGC\,3079 ($0.2 - 2$
  keV) by a self-consistent NEI outflow
  model (solid line; Breitschwerdt et al. 2003); the lower panel shows the fit residuals.
  \textbf{Right:}
  PSPC R4 ($520 - 690$ eV) and R5 ($700 - 900$ eV) band ratio in a
  disk parallel stripe ($\Delta z = 2$ kpc) along the minor axis
  of NGC\,253 (solid histogram) compared to OVII /FeXVII
  flux ratio averaged over $625 - 675$ eV and $805 - 865$ eV,
  respectively (dashed histogram), derived from the best fit halo
  model, according to self-consistent NEI calculations (cf.~Fig.~\ref{n253-img}).
  }
\end{figure}
The data are shown as crosses (error bars) and clearly exhibit two
humps, which we identify as due to oxygen (at around 0.5 keV) and
iron L shell line ($0.8 - 0.9$ keV) complexes. This is
corroborated by running again an outflow simulation, generating a
self-consistent NEI spectrum and subjecting it to the same
procedure as before (i.e. folding through the EPIC pn instrumental
response). The best model fit (with a reduced $\chi^2 \approx
1.2$) requires a temperature of $T = 3.56 \times 10^6$ K and a
density of $n= 5 \times 10^{-3}\, {\rm cm}^{-3}$ at the base of
the halo. The outflow velocity there is about 220 km/s, reaching
about 450 km/s at $z=50$ kpc. The higher initial temperature as
compared to NGC\,253 reflects the more violent outflow in this
case and is necessary for producing the iron bump. We find again
that no single RS model can fit the spectrum.
The discrepancy between the cooling rates in CIE and NEI can be
quite substantial at certain temperatures and thus have a
pronounced effect on the outflow dynamics, as can be seen by
comparing the respective cooling curves in Fig.~\ref{n253-spec}:
delayed recombination can indeed reduce the cooling rate.

To go one step further, if photon statistics of the observational
data permits, one can analyze the spectral behaviour of the plasma
in disk parallel stripes of widths of $\Delta z \sim 2$ kpc using
a ``sliding window'' technique (s.~Fig.~\ref{sbgal_fit}). The idea
is to search for ``diagnostic ions'' of starburst driven outflows.
The double hump in Fig.~\ref{sbgal_fit} suggests OVII and FeXVII
as spectral signatures that should vary in strength as the outflow
cools down adiabatically and radiatively. We have therefore
compared the PSPC R4 ($520 - 690$ eV) to R5 ($700 - 900$ eV) band
ratios along the minor axis to an averaged OVII ($625 - 675$) eV
to FeXVII ($805 - 865$ eV) ratio of the model calculations.
Although the latter ones have arbitrarily high ``spectral
resolution'', we have averaged over an energy range free of other
spectral features; hence the energy range is smaller than in the
corresponding ROSAT PSPC bands, and the absolute fluxes do not
match. However, it can be seen that the flux ratio of these
characteristic ions have pronounced peak at about 1.5 kpc from the
disk both for the observed and the model spectra. This gives a
good hint that the time-dependent evolution of these ion species
in the outflow should be basically correct. More stringent
conclusions have to be deferred to an analysis of higher
resolution spectra.

\section{Conclusions}
\label{conc}
The discrepancy between equilibrium plasma models and
observational data in the EUV and soft X-rays is commonly felt as
a nuisance. There exists e.g.\ a plethora of high spectral
resolution data for the Local Bubble. It is indeed embarrassing
that presently neither SNR nor SB models, with CIE or simple NEI
cooling can reproduce all these data satisfactorily. However, I
would like to stress a positive aspect of this problem: the
spectral fingerprints obtained contain far more information than
our simple models can handle. Although being intrinsically more
complicated, the advantage of NEI models is to connect the
spectrum of a plasma at any given time to its history and
therefore to help us to learn something about the plasma's
excitation conditions and hence its origin. In case of the Local
Bubble it seems that more sophisticated 3D simulations including
NEI cooling, as undertaken currently (Avillez \& Breitschwerdt,
2003), should lead to an improvement.

Meanwhile it is worth focusing also on somewhat simpler
astrophysical problems such as galactic outflows, which are well
observable in edge-on galaxies. I have shown that self-consistent
modelling can explain a number of problems: (i) the extension of
soft X-ray halos, (ii) their spectra (although better photon
statistics and higher spectral resolution are desirable), (iii)
highly subsolar abundances in starburst galaxies are artifacts of
the fit procedure, (iv) depending on the origin of the plasma
there exist diagnostic ions that contribute to the emission in a
characteristic fashion during the evolution of the outflow (along
the minor axis in edge-on galaxies). Further studies will enable
us to eventually connect the spectral behaviour to the star
formation activity.

%\acknowledgements The XMM-Newton project is supported by the
%Bundesministerium für Bildung und Forschung/Deutsches Zentrum
%f\"ur Luft- und Raumfahrt (BMBF/DLR), the Max-Planck-Gesellschaft
%and the Heidenhain-Stiftung.

\end{article}
\end{document}